# Hardware-Aware Data and Instruction Mapping for AI Tasks: Balancing Parallelism, I/O and Memory Tradeoffs

Md Rownak Hossain Chowdhury, Mostafizur Rahman
Division of Energy, Matters, and Systems, University of Missouri-Kansas City (UMKC)
Kansas City, MO, US
{rhctmc, rahmanmo}@umkc.edu

*Abstract*— We introduce a mapping framework for deep learning inference that takes advantage of predictable neural network behavior to plan both computation and communication ahead of time. The framework generates a unified stream of instructions and data, enabling the hardware to execute operations and route information on its own, without frequent involvement from the host and with minimal off-chip memory use. This naturally reduces reliance on I/O, off-chip memory, and host control. By leveraging fine-grained message passing on a programmable, message-based compute architecture, the framework keeps data movement local and coordinates computation across the array using techniques such as stationary-weight reuse, in-array multicasting, and staged reductions. Applied to VGG-19, the framework sustains high utilization (88 to 92 percent), with over 97 percent of messages generated internally and nearly 89 percent of time consumed on-chip transfers. Computation throughput scales beyond 1 TFLOP/s on larger arrays, while traffic reductions from reuse and local aggregation reach up to 100 MB per layer. Overall, the results highlight the effectiveness of streaming-based computation and show how our mapper enables this execution style by tightly coordinating data and instruction flow across the hardware.

## I. INTRODUCTION

Domain-specific AI accelerators achieve high compute density by specializing data paths and exposing spatial parallelism **[1]**, **[2]**, **[3]**, **[4]**, **[5]**. However, in real-world deployments, the overall performance of deep learning inference is often limited not by arithmetic capability, but by the challenges of software mapping, I/O orchestration, and memory access patterns **[6]**, **[7]**, **[8]**, **[9]**. Transferring large volumes of weights and activations across PCIe and off-chip DRAM introduces latency and bandwidth bottlenecks that significantly degrade end-to-end performance **[10]**, **[11]**, **[12]**. This results in a persistent compute-communication gap: while the chip may offer teraflops of raw compute, off-chip interfaces typically provide only tens of gigabytes per second, constraining throughput **[13]**, **[14]**.

Dataflow-centric accelerators aim to reduce memory and I/O bottlenecks by reusing data on-chip and broadcasting inputs to minimize repeated fetches **[15]**, **[16]**, **[17]**. Coarse-grained reconfigurable arrays (CGRAs) extend this approach by mapping computations onto a grid of programmable tiles connected via on-chip networks **[18]**, **[19]**, **[20]**. While both architectures seek to exploit locality and spatial parallelism, they share a common reliance on host-managed DMA and statically compiled micro-kernels **[21]**, **[22]**, **[23]**. In typical deployments, activations and weights are loaded into on-chip SRAM, partial results are written back, and layers are processed in tiles. Transitioning across layers or handling boundaries (e.g., padding or strides) requires flushing state and reprogramming the array, which breaks opportunities for reuse **[24]**, **[25]**. CGRAs offer more flexibility, but this comes at the cost of complex placement, routing, and tightly synchronized per-tile instruction schedules that consume instruction bandwidth and local storage **[26]**. These rigid execution models limit the ability to adapt data movement and control flow at runtime, leaving performance bottlenecked by the host's ability to feed and manage the array. While architectural techniques like larger SRAMs **[27]**, **[28]**, compression **[29]**, **[30]** and prefetching **[31]**, or faster DRAM **[32]** and interconnects **[33]** help ease bandwidth pressure, they do not address the underlying constraint: a host-driven execution model that repeatedly moves both control and data off-chip, constraining scalability and efficiency.

In our work, we take the view that deep-learning inference is structured enough to shift control away from the host. The model defines tensor shapes, layer order, and kernel connectivity, while the accelerator's topology and dataflow are fixed at design time. This makes both the spatial placement of computation and the timing of communication predictable before execution. We compile this spatial–temporal plan into a message stream that carries both data and control, allowing execution to unfold as a self-driven process: data flows along pre-defined routes, compute units activate in place, and each stage triggers the next without host intervention. This design eliminates load-execute-store boundaries, reduces off-chip traffic, and smooths out I/O latency, enabling the fabric to sustain near-continuous operation once primed.

We realize this idea via a message-driven execution model where a single 64-bit packet co-packs the opcode, present/next addresses, and a 32-bit operand (weight, activation, or partial sum). Processing nodes consume a packet, performs the operation, and rewrites it with the follow-on opcode and destination, fusing compute and communication into one pipeline. Once primed, the array streams to completion without relaunching driver and avoiding repeated off-chip fetch/store cycles. Section *III* details packet format, addressing, and scheduling. Following are the core contributions of this paper:

- We present a message-based execution model that co-encodes data and control, enabling self-sequencing on the fabric and reducing reliance on off-chip memory and I/O.

- We demonstrate that over 97% of messages during VGG-19 inference are generated on-chip, highlighting strong decoupling from both the host and DRAM.
- We provide a detailed evaluation of resource utilization, data movement, reuse, and throughput, showing 88–92% utilization, over 1 TFLOP/s on convolution layers, reduced latency at larger array sizes, and high spatial/temporal reuse via on-chip multicast and staged reductions. System-level throughput (KIPS) remains largely stable across PCIe and DRAM bandwidth variations.

The remainder of this manuscript is structured as follows. Section *II* summarizes the MAVeC accelerator architecture. Section *III* formalizes the messaging-based data-and-instruction streaming and presents a compact case study. Section *IV* evaluates MAVeC on VGG-19. Finally, section *V* concludes with key findings.

## II. MAVEC ACCELERATOR

MAVeC follows System-on-Chip (SoC) organization (*Figure 1*, left), using a PCIe-mediated system bus to bridge the host CPU, SSD, and off-chip DRAM via a dedicated controller. The host writes 64-bit instructions and 32-bit weights into DRAM. The controller then streams these packets into on-chip Tile buffers (L2). From there, data cascades through SiteM banks (L1) into SiteO registers (L0), staging operands and intermediate results for execution. MAVeC embodies a self-programming ASIC that reconfigures on the fly for each neural-network layer without host intervention, aligning the pipeline to each layer's work. Layer-specific directives reside in the 64-bit instruction fields, so every packet carries both its operation and routing fields. As messages traverse the array, each SiteO executes its opcode, stages or evicts local data, and forwards an updated packet to its downstream neighbor. The stream naturally reconfigures at every layer boundary, sustaining high utilization and minimizing off-chip access across convolutional, pooling, and fully connected stages.

MAVeC uses a unified 64-bit message (*Figure 2*) that encapsulates address, opcode and data to perform Computation and routing. The first 4-bit field specifies the Present Opcode, followed by a 12-bit Present Address that targets a specific SiteO. A 32-bit payload carries either a filter index during initialization or a partial-sum/data value during execution. The final 4-bit Next Opcode and 12-bit Next Address embed the downstream operation and destination for forwarding. *Table 1* lists MAVeC's 13-opcode ISA: **Prog** seeds weight streams, while **UPDATE**, **A_ADD(S)**, **A_SUB(S)**, **A_MUL(S)**, **A_DIV(S)**, **Av_ADD**, **RELU**, and **CMP** perform arithmetic or activation and drive message propagation. This co-packed format delivers decentralized, pipelined execution and alleviates off-chip memory and I/O bottlenecks.

**Table 1. MAVeC Instruction Set Architecture (ISA)**

| Instruction | Opcode | Task |
|---|---|---|
| Prog | 0001 | Store weights and routing data |
| UPDATE | 1101 | Update SiteO with incoming data |
| A_ADD | 0100 | Update SiteO after addition |
| A_ADDS | 0111 | Stream addition result |
| A_SUB | 0101 | Update SiteO after subtraction |
| A_SUBS | 1000 | Stream subtraction result |
| A_MUL | 0010 | Update SiteO after multiplication |
| A_MULS | 1001 | Stream multiplication result |
| A_DIV | 0110 | Update SiteO after division |
| A_DIVS | 1010 | Stream division result |
| Av_ADD | 1011 | Update SiteO after averaging |
| RELU | 0011 | ReLU activation operation |
| CMP | 1100 | Update SiteO after comparison |

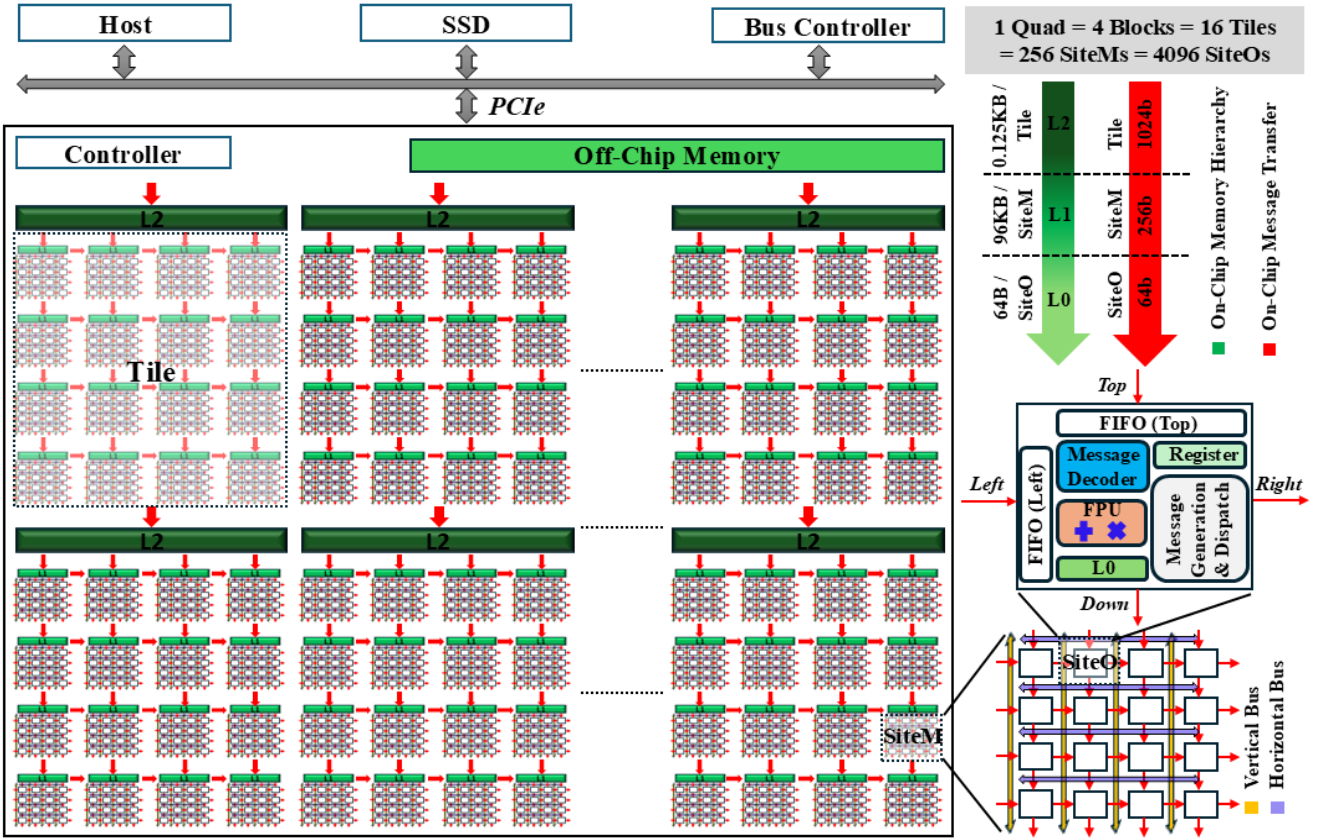

**Figure 1. MAVeC's integrated streaming fabric within SoC.** **(Left)** A PCIe-mediated system bus linking the host CPU, SSD, and off-chip DRAM to a dedicated controller. **(Right)** On-chip L2/L1/L0 buffers span a SiteOs → SiteMs → Tiles → Blocks → Quads hierarchy; red arrows trace message streams, yellow/blue buses enable multicast and reduction, and the inset details the SiteO/SiteM microarchitecture.

MAVeC's compute fabric is organized into a five-level hierarchy (SiteOs → SiteMs → Tiles → Blocks → Quads); yielding 4,096 SiteOs per Quad (*Figure 1*, right). Each SiteO (*Figure 1*, inset) contains a 32-bit IEEE-754 FPU, an instruction decoder, two input FIFOs (Top and Left), and an 8-word local buffer for weight storage. Incoming messages arrive via the Top or Left FIFO, compare their 12-bit destination field to the local address, and either execute their opcode on the match or forward the packet to the bottom or right FIFO in the same cycle, asserting a flow-control stop only if a target FIFO is full. Within each 4x4 SiteM, dedicated vertical (yellow) and horizontal (blue) buses (*Figure 1*, inset) enable parallel multicast for input/output broadcasts from SiteOs and reduction for partial-sum aggregation. This bus-based routing extends through Tiles, Blocks, and Quads to deliver low-latency, fine-grained communication to sustain high utilization across the accelerator [34], [35].

To enable in-fabric streaming, MAVeC implements a three-tier on-chip memory hierarchy: L2 Tile buffers, L1 SiteM banks, and L0 SiteO registers, as shown by the green arrows in *Figure 1*. Each L2 Tile buffer provides 0.125 KB of storage and connects to the PCIe controller via sixteen 1024-bit links to absorb off-chip bursts. At L1, each SiteM bank allocates 96 KB to hold incoming messages and intermediate results independently, linking to Tiles with sixty-four 256-bit buses and to neighboring SiteMs with 192 additional 256-bit channels. At L0, each SiteO register file stores 64B of weights and local data alongside its FPU, enabling single-cycle operand access. Messages flow from DRAM to L2, then to L1, and finally to L0, as illustrated by the red arrows in *Figure 1*. These on-chip memory levels provide 24.5 MB per Quad (under 100 MB per Core) of near-compute storage. Combined with MAVeC's messaging-based execution model, this hierarchy reduces off-chip memory traffic [36].

MAVeC diverges from existing microarchitectures in multiple aspects that impact end-to-end system performance. First, it relocates sequencing into the fabric instead of relying on host-paced kernels or central command stream (as in systolic/dataflow arrays) or long per-tile instruction traces (as in CGRAs). MAVeC binds the task and its routing target to the same packet, so execution advances hop-by-hop without a phase break. Second, it redefines the compute primitive exposed to the array. Rather than the MAC-only data-path tailored to lowered GEMMs, each SiteO executes a small arithmetic/activation operation while forwarding next operation in flight, which preserves convolution semantics and avoids wholesale tensor lowering. Third, it replaces global NoC waves and DMA epochs with hierarchical multicast-and-reduce buses and on-chip memory, letting data propagate naturally and partials collapse locally, so layer boundaries become soft handoffs rather than drain-and-reload points. Consequently, the system operates as a single resident pipeline in which packets carry operands and next-step directives, intermediates need not reappear off chip, and the fabric reconfigures itself at layer granularity.

### III. HARDWARE-CENTRIC AUTONOMOUS DATA & INSTRUCTION STREAMING

In this section, we first show how packets are generated on-chip, then outline the message-driven dataflow, how addresses are allocated deterministically through the array, and how tasks are scheduled without host control. We close with a compact end-to-end walkthrough that ties these components together.

#### A. On-Chip Message Generation

MAVeC's computation begins with a 64-bit message from the host (*Figure 2*), composed of a 4-bit present opcode, a 12-bit present address, a 32-bit payload, a 4-bit next opcode, and a 12-bit next address. The host issues two message types: **Prog** for filter-weight and routing setup, and Compute (**A_MULS**) for arithmetic. In **Prog** messages, active columns (rainbow) carry filter weights, while interleaved reserved columns hold zeros (gray). Compute messages use the 32-bit payload to transport activations or partial sums. For Compute messages, the upper 16 bits (Next fields) are zero for 1x1 convolutions and fully connected layers, whereas for larger kernels they encode the workload pattern. The workload pattern carries designated flags like $T_{Stream}$ for inter-tile (Group$_{Next}$) transfers, **Shift** for intra-tile (SiteO$_{Next}$) routing, and **Identity** bit for skip connections (e.g., ResNet shortcuts), with unused bits left available for future extensions.

Upon receiving a compute message, each SiteO follows a four-step process to generate intermediate message, as shown in the bottom portion of *Figure 2*. First, it decodes the present opcode, present address, and 32-bit payload from the incoming compute messages. Second, it executes a specified operation such as multiplication (**FxD**) and produces a 32-bit result. Third, it reads the existing next opcode, next address, and workload-pattern bits from incoming messages and assigns them into the new present opcode and present address fields. Finally, it generates the intermediate messages by concatenating updated present fields, the 32-bit result, and pattern bits. This newly emitted message thus carries both the result of the local computation and the exact routing and workload directives needed by the next SiteO. Next, we describe how generated messages traverse the hardware array via a fixed address map.

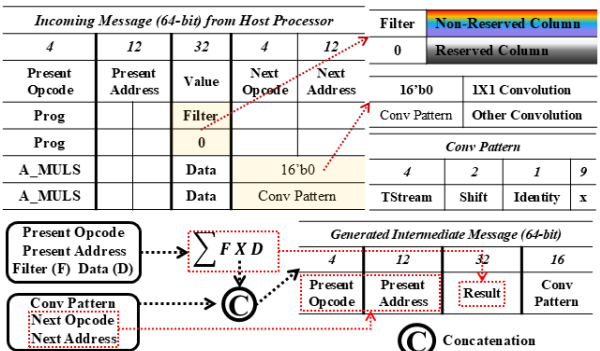

**Figure 2. On-Chip Message Generation.** Incoming 64-bit messages from host carry a Present Opcode, Present Address, 32-bit payload (filter weights or data), Next Opcode, and Next Address; at each SiteO, these fields are unpacked, the specified operation is performed, and the result is concatenated with updated Present fields and workload-pattern bits into a new message.

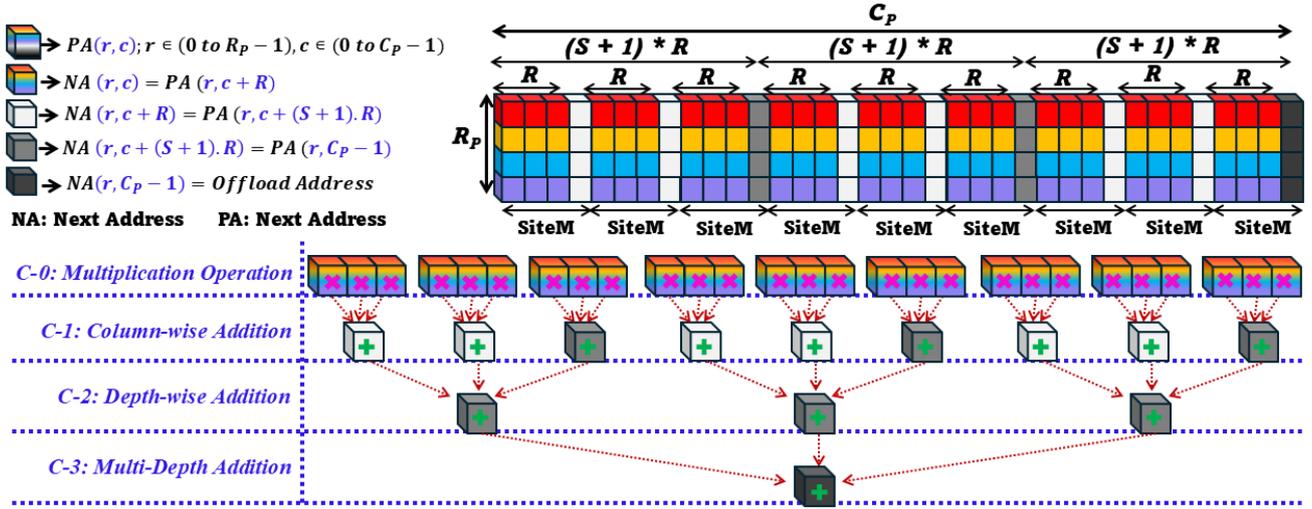

Figure 3. Deterministic On-Chip Routing for Multi-Stage Reduction. SiteO arrays split into active (C-0) and reserved columns (C-1, C-2, C-3) for successive spatial reduction. Intermediate messages route deterministically from C-0 → C-1 → C-2 → C-3 and offload to L1 on-chip memory for subsequent stages.

### B. Address Allocation

Intermediate messages traverse the SiteO array ($R_P$ x $C_P$) in a deterministic four-stage pipeline during each image block (IB)-filter fold (FF) interaction, as shown in *Figure 3*. The SiteO columns are divided into an active set (C-0) for initial multiplications and three reserved sets (C-1, C-2, C-3) for successive reductions. Initially, C-0 columns groups R adjacent columns, each multiplying its local weight by the incoming data in parallel. When all R products complete, they broadcast over the horizontal bus into the C-1 columns, which perform column-wise sums. This routing sets fixed next address (NA (r, c) = PA (r, c+R)) of C-0 columns to move multiplication results to C-1 columns.

Next, each C-1 column forwards its summed result over the horizontal bus into the C-2 columns using the offset NA (r, c+R) = PA (r, c+(S+1)*R), enabling depth-wise accumulation. Finally, all results of C-2 columns are forwarded to offload columns (C-3) at PA (r, $C_P$-1) to aggregate multiple depths. Packets arriving at PA (r, $C_P$−1) then offload their fully reduced scalars into L1 SiteM bank for subsequent addition, pooling or the next convolutional layer. Thus, MAVeC orchestrates data following a deterministic routing pattern in four stages (C0 → C1 → C2 → C3) using the messaging-based execution model without host intervention and costly off-chip fetches.

### C. Task Allocation & Message Scheduling

In the preceding subsection, we described how MAVeC routes generated messages through a fixed, four-stage address map (C0 → C1 → C2 → C3). Here, we focus on deterministic task allocation, showing how each SiteO's incoming message deterministically produces an outgoing message that both completes local operation and schedules the next stage or the next layer. *Table 2* lists the sent (left) and generated (right) messages across convolution, activation, and pooling, so the pipeline - from programming through reduction and handoff to the next layer - runs autonomously.

At the beginning of a convolution pass, every SiteO receives setup messages to prepare for streaming (*Table 2, entries 1-7*). During programming (Prog), C-0 SiteOs are loaded with filter weights (*entry 1*), while C-1/C-2/C-3 SiteOs are seeded with zero to clear their accumulators (*entries 3-7*). The next opcode in each Prog message pre-arms the reduction behavior. For C-0 through C-2 it is A_ADDS (add-and-stream): once products arrive, C-1 performs the column-wise sum and forwards to C-2, which performs the depth-wise sum and forwards to C-3. For C-3, the next opcode depends on the Image Fold (IF)-Filter Fold (FF) interaction sequence: UPDATE for the first fold (*entry 5*) to initialize the offload address (OA) with the initial multi-depth summation, A_ADDS for intermediate folds (*entry 7*) to keep accumulating into OA, and A_ADD for the last fold (*entry 6*) to finish the accumulation and hold the result.

Immediately after programming, A_MULS compute messages are multicast to all C-0 SiteOs (*entry 2*), each carrying an image data and the convolution-pattern bits. On reception, C-0 SiteOs multiply programmed weight (F) by the image value (In) and emit A_ADDS@C-1 with value (FxIn) for column-wise reduction (*entry 1*). For kernels larger than 1x1, the same event also creates two forwarding packets driven by the pattern flags: Shift flag produces an A_MULS (Present Address = SiteO_Next), supplying the image data for the next in-tile shift (*entry 2.1*); T_Stream produces another A_MULS (Present Address = Group_Next), pre-staging the data for the next tile group (*entry 2.2*). The remainder proceeds as in the generated side of *Table 2*: C-1 accumulates R products (*entry 3*), C-2 accumulates across S (*entry 4*), and C-3 applies UPDATE/A_ADD/A_ADDS for first/last/rest folds to emit messages whose Present Addresses is OA (*entries 5-7*). These OA-addressed packets are written into the L1 on-chip buffer and later fetched at OA by subsequent stages for further accumulation.

**Table 2. Deterministic Task Allocation and Message Schedule across Streaming Pipeline**

| Sl. No. | ID$_{IB}$-ID$_{FF}$/ Next Layer# | SENT MESSAGES | | | | | GENERATED MESSAGES | | | | |
|---|---|---|---|---|---|---|---|---|---|---|---|
| | | Present Opcode | Present Address | Value | Next Opcode | Next Address | Present Opcode | Present Address | Value | Next Opcode | Next Address |
| 1 | All | Prog | C0 | F | A_ADDS | C1 | A_ADDS | C1 | $F * In$ | – | – |
| 2.1 | | A_MULS | C0 | In | Conv Pattern | | A_MULS | SiteO$_{Next}$ | In | – | – |
| 2.2 | | | | | | | A_MULS | Group$_{Next}$ | In | – | – |
| 3 | | Prog | C1 | 0 | A_ADDS | C2 | A_ADDS | C2 | $\sum_R F \times In$ | – | – |
| 4 | | Prog | C2 | 0 | A_ADDS | C3 | A_ADDS | C3 | $\sum_S F \times In$ | – | – |
| 5 | 1$^{st}$ | Prog | C3 | 0 | UPDATE | OA | UPDATE | OA | $\sum_{C=1:X} F \times In$ | – | – |
| 6 | Last | Prog | C3 | 0 | A_ADD | OA | A_ADD | OA | $\sum_{C=Y:C} F \times In$ | – | – |
| 7 | Rest | Prog | C3 | 0 | A_ADDS | OA | A_ADDS | OA | $\sum_{C=X:Y} F \times In$ | – | – |
| 8 | Conv# (non-1X1) | ReLU | OA | 0 | A_MULS | C0 | A_MULS | C0 | $R(\sum_{All-C} F \times In)$ | Conv Pattern | |
| 9 | Conv# (1X1) | ReLU | OA | 0 | A_MULS | C0 | A_MULS | C0 | $R(\sum_{All-C} F \times In)$ | 16'b0 | |
| 10 | FC# | ReLU | OA | 0 | A_MULS | C0 | A_MULS | C0 | $R(\sum_{All-C} F \times In)$ | 16'b0 | |
| 11 | Max Pool# | ReLU | OA | 0 | CMP | C0 | CMP | C0 | $R(\sum_{All-C} F \times In)$ | 16'b0 | |

Once all weight and image folds interactions for a layer complete, partial sums are fetched from L1 at their OA SiteO coordinates and accumulated until all-depth sum is available. MAVeC then issues activation messages (*Table 2, entries 8-11*) that apply **ReLU** to the sum {$R(\sum_{All-C} F \times In)$} and format the input stream for the next layer. If the next layer is a regular convolution (>1x1), MAVeC generates **A_MULS@C-0** with ReLU output and the next layer's Conv-Pattern (*entry 8*). If the next layer is pointwise convolution (1x1) or fully connected (FC); it generates **A_MULS@C-0** with the same ReLU output and a zero pattern (**16'b0**), since no intra- or inter-tile shifts are required (*entry 9-10*). When the next layer is max pooling, it generates **CMP@C-0** to seed the comparison chain (*entry 11*). All these messages are written back to L1 and staged at OA so the next layer can start streaming without host intervention.

### D. Message-Driven Dataflow

The canonical seven-dimensional convolution loop nest over number of filters (**N$_F$**), channels (**C**), filter height (**R**), filter width (**S**), batch elements (**N**), and output positions (**P**, **Q**) can be mapped onto spatial fabric, as described in [37]. Here, we provide a concise overview. First, we flatten the 4D filter-tensor into a 2D matrix using depth-major traversal (**C** before **R** and **S**) and column-wise unrolling of each **RxS** kernel, inserting one reserved column after every row to enable reduction. We then slice this matrix into Filter Folds (**FF**) according to the available SiteO array's (**R$_P$** rows and **C$_P$** columns). Concurrently, we partition the input-tensor into Image Blocks (**IB**) that match each **FF**'s channel depth. From each **IB**, we generate (**PxN**) Image Folds (**IF**) - P folds per image over **N** images - by sliding a width-**S** window across the spatial dimension, so each **IF** contains **S** columns from every matched channel. Each **IF** then advances through **Q** shifts, moving by convolution stride at each step to cover all output positions. To fit the Filter Fold layout and prevent redundant fetches, we reverse the order of columns within each **IF** and drop any columns that repeat across consecutive folds.

Execution unfolds entirely via MAVeC's 64-bit message stream rather than traditional load-execute-store cycles. A Filter Fold is first programmed into SiteO registers (L0), keeping its weights stationary to maximize temporal reuse. Thereafter, each Image Fold is injected: activations are multicast across rows via vertical buses, enabling parallel element-wise multiplications that generate 32-bit partial sums. These partial sums propagate through the spatial array and undergo hierarchical reduction - first across each column, then within depth slices, and finally across multiple depths - to yield a single output per position. Once reduction is completed, the image fold advances by the configured convolution stride, and the pipeline repeats until all output columns are produced.

### E. Case Study: End-to-End Streaming Walkthrough

In this subsection, we illustrate MAVeC's message-driven data-and-instruction stream on a compact layer - input **4x4x4**, filter **3x3x4**, **8** filters, stride **1**, pad **1**, ReLU - producing **4x4x8** (*Table 3(A)*). Using the mechanisms introduced earlier (folding, on-chip routing, message generation, and scheduling), we show exactly how the layer executes end-to-end on the fabric. The host-side mapper first

**Table 3. A)** Specifications of example workload **B)** Mapping constructs (FF, IB, IF) generated by host-side mapper

**A)**

| Input (X, Y, C) | Input ($X_{Pad}$, $Y_{Pad}$, C) | Filter (R, S, C) | #Filters ($N_F$) | Stride | Pad | Activation | Output (P, Q, $N_F$) |
|---|---|---|---|---|---|---|---|
| 4 X 4 X 4 | 6 X 6 X 4 | 3 X 3 X 4 | 8 | 1 | 1 | ReLU | 4 X 4 X 8 |

**B)**

| $ID_{FF}$ | $FF_{Shape}$ | $ID_{IB}$ | $ID_{IF}$ | $Col_{IF}$ | (C) @ $C_P$ | ($N_F$) @ $R_P$ | $ID_{PS}$ | $PS_{Shape}$ | $O(N_F)^C_{Index}$ |
|---|---|---|---|---|---|---|---|---|---|
| #1 | 4 X 24 | #1 | #1 | {0, 1, 2} | 0, 1 | 0 to 3 | #1 | 4 X 16 | $O(0:3)^{0:3}_{0:15}$ |
| | | | #2 | {3} | | | | | |
| | | | #3 | {4} | | | | | |
| | | | #4 | {5} | | | | | |
| #2 | 4 X 24 | #2 | #1 | {0, 1, 2} | 2, 3 | | #2 | 4 X 16 | |
| | | | #2 | {3} | | | | | |
| | | | #3 | {4} | | | | | |
| | | | #4 | {5} | | | | | |
| #3 | 4 X 24 | #3 | #1 | {0, 1, 2} | 0, 1 | 4 to 7 | #3 | 4 X 16 | $O(4:7)^{0:3}_{0:15}$ |
| | | | #2 | {3} | | | | | |
| | | | #3 | {4} | | | | | |
| | | | #4 | {5} | | | | | |
| #4 | 4 X 24 | #4 | #1 | {0, 1, 2} | 2, 3 | | #4 | 4 X 16 | |
| | | | #2 | {3} | | | | | |
| | | | #3 | {4} | | | | | |
| | | | #4 | {5} | | | | | |

targets **4x24** SiteO array and reshapes the layer into the hardware constructs - Filter Fold (**FF**), Image Block (**IB**), and Image Fold (**IF**) - as summarized in **Table 3(B)**. Each block of rows corresponds to one **FF** of shape **4x24**. The ($N_F$)@$R_P$ column indicates the filter rows placed on the array (filters **0-3** for **FF#1-2**, and **4-7** for **FF#3-4**). The (C)@$C_P$ column records the channel group mapped across columns - {**0,1**} for **FFs #1** and **#3** and {**2,3**} for **FFs #2** and **#4**. For each **FF**'s matching **IB**, the mapper generates four Image Folds (**IF #1** to **#4**). $Col_{IF}$ lists unique input columns per fold ({**0, 1, 2**}, then {**3**}, {**4**}, {**5**}) for assigned depths. For every **FF-IB** pair, $ID_{PS}$ and $PS_{Shape}$ identify the partial-sum tile produced on chip (**4x16** here), while $O(N_F)^C_{Index}$ denotes the contiguous

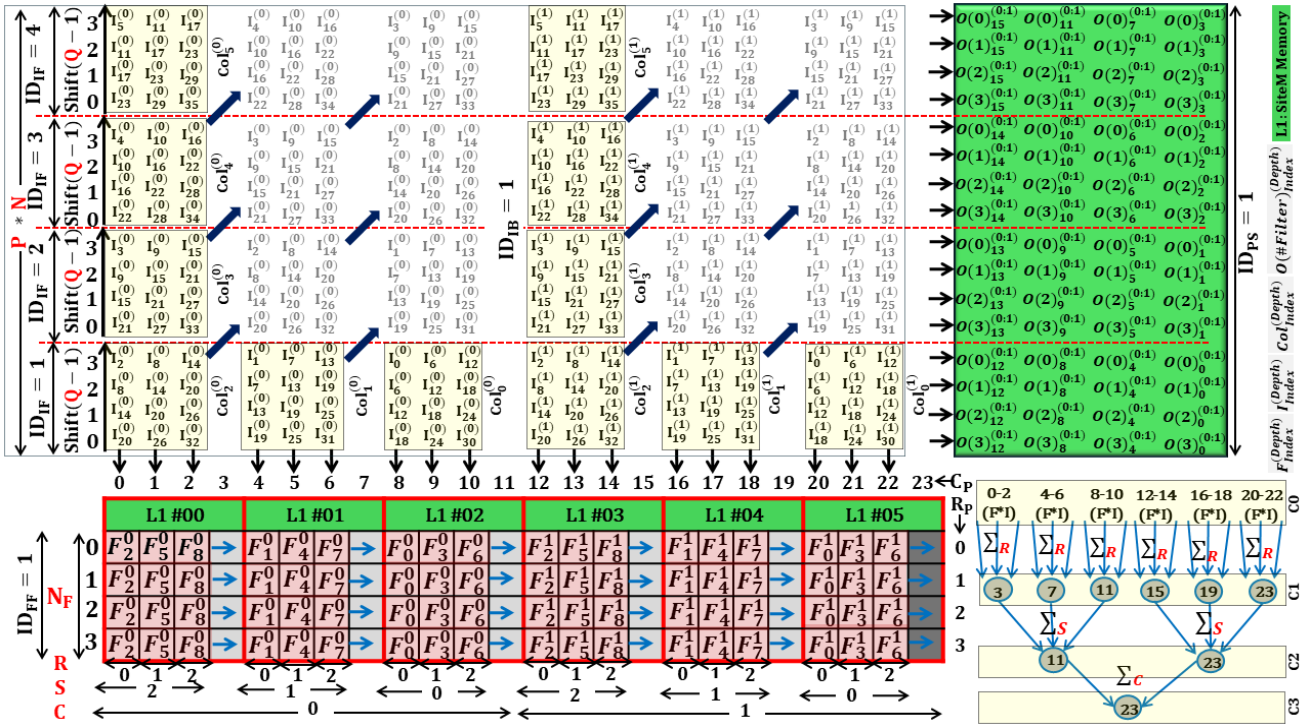

**Figure 4. One FF-IB pass under the message-driven pipeline.** FF#1 is programmed on the 4x24 SiteO array. IF#1 is multicast to perform multiplication (FxIn) parallelly. Multiplication result flow through hierarchical reduction ($\Sigma_R \rightarrow \Sigma_S \rightarrow \Sigma_C$). IF#2 to IF#4 fetch only new image columns {3}, {4}, {5} while overlapping columns are forwarded on-chip indicated by blue arrows. The resulting output ($O(N_F)^C_{Index}$) messages are then written to L1 on-chip memory.

slice of the output tensor (filter range and output indices) covered by that tile.

*Figure 4* shows one **FF-IB** pass under the messaging pipeline. **FF#1** (filters **0-3** over channels **{0,1}**) is programmed into the **C-0** columns (**0-2**, **4-6**, **8-10**, **12-14**, **16-18**, **20-22**). Each **C-0** column holds filter values in depth-major ($F^0 \rightarrow F^1$) column-reversed order ($F_2, F_5, F_8 \rightarrow F_1, F_4, F_7 \rightarrow F_0, F_3, F_6$). In that fashion, four distinct filters ($N_F \rightarrow 0-3$) are distributed in four hardware rows ($R_P \rightarrow 0-3$). Then, **IF#1** injects columns **{Col$_2$, Col$_1$, Col$_0$}** for channels {0, 1}. Here, aligned pixels ($I_{Index}^{(Depth)}$) are multicast across rows, and each **C-0** multiplies its local weight ($F_{Index}^{(Depth)}$) by the incoming value ($I_{Index}^{(Depth)}$) parallelly. The products fan out to **C-1** (3, 7, 11, 15, 19, 23) for column-wise reduction ($\sum_R$); those sums advance by the fixed **(S+1).R** offset to **C-2** (11, 23) for depth-wise reduction ($\sum_S$); they then jump to **C-3** (**C$_P$-1**) for multi-depth reductions ($\sum_C$), and the results are written to L1.

After that, each image column shifts right **Q-1** times by convolution stride as depicted in the left portion of *Figure 4*. This pipeline (shift → multicast → multiply → reduction) continues until one output column over all filters is generated. Then, **IF#2-IF#4** inject **Col$_3$**, **Col$_4$**, and **Col$_5$** for channels {0, 1}; overlapping columns are forwarded laterally (blue arrows), so only the new column is fetched each time. After these four shifts the four output columns for filters **0-3** over channel {0,1} are complete (**PS#1**). Next, the array programs **FF#2** (filters **0–3** over {2,3}) and streams **IB#2** to produce **PS#2**. The rows are then reprogrammed with **FF#3** (filters **4-7** over {0,1}) to produce **PS#3**, and finally with **FF#4** (filters **4-7** over {2,3}) to produce **PS#4**. At the end of these four passes, **PS#1-PS#4** messages reside in L1 with present address OA, with shapes and indices matching *Table 3(B)*. Thus far, the pipeline's consumed and produced messages correspond exactly to *Table 2 (entries 1–7)*.

*Figure 5* depicts the final depth merge and layer hand-off. Once all **FF–IB** passes complete, MAVeC reads the partial-sum tiles (**PS#1-PS#4**) from L1 at their OA coordinates on the SiteO grid and executes a streamed merge to produce and stage the next-layer inputs. For the first filter band (**N$_F$ 0-3**), **PS#1** is streamed to OA with **UPDATE** to initialize the accumulator, followed by **PS#2** with **A_ADD** to form the all-depth sum. ReLU is applied in place; the activations are then encoded as **A_MULS** packets - parameterized with the next layer's Conv-Pattern (or **16′b0** for 1x1/FC) - and written to L1 as the next-layer inputs. The second band (**N$_F$ 4–7**) follows the same sequence: **PS#3** with **UPDATE**, then **PS#4** with **A_ADD**, in-place ReLU, and emission of **A_MULS** packets to L1. Deterministic task allocation and message scheduling thus performs both the merge and the hand-off to the subsequent layer without host intervention.

## IV. PERFORMANCE ANALYSIS

### A. Evaluation Methodology

We evaluate MAVeC - a spatially programmable, message-driven ASIC targeting the TSMC 28 nm node - at 1 GHz with FP32 arithmetic. The baseline platform for evaluation assumes a PCIe Gen6x16 host link and DDR7 as the off-chip memory. The workload is the VGG-19 convolutional stack (batch = 1) summarized in *Table 4*. To assess I/O sensitivity, we sweep PCIe configurations (*Table 5(A)*) and DRAM families (*Table 5(B)*). Each layer is compiled under the message-driven execution model elaborated in Section ***III*** and executed on three array sizes (16x16, 32x32, and 64x64 SiteOs) to study scaling. Measurements are obtained with a cycle-accurate, message-level simulator leveraging analytic models in **[36], [37]**. For VGG-19 network, the simulator reports: (i) message count (host-injected weight/image vs. on-chip–generated messages); (ii) a cycle breakdown by execution phase (message transfer, operation, host to off-chip memory, weight load); (iii) average SiteO utilization; (iv) total latency in clock cycles; and (v) compute throughput in GFLOP/s. To highlight data locality, we also quantify temporal reuse, spatial reuse, spatial reduction (all in MB), where temporal reuse captures bytes re-used from stationary operands, spatial reuse captures bytes saved by in-array multicast, and spatial reduction reflects bytes elided by reduction chain.

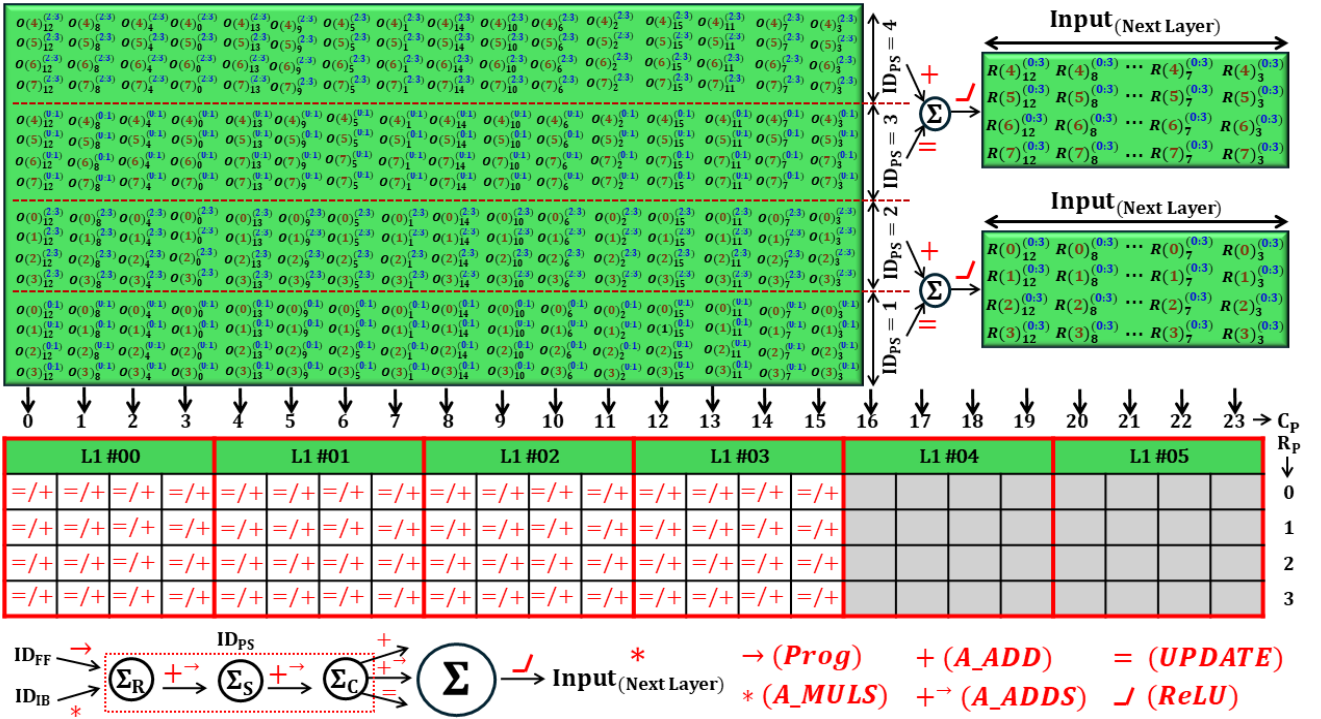

**Figure 5. Final on-chip merge and hand-off to the next layer.** Partial sum folds residing at L1 are streamed per filter band (N$_F$ 0-3, then 4-7). PS#1/PS#3 with UPDATE seed the accumulator; PS#2/PS#4 with A_ADD complete the all-depth sum. ReLU is applied in place, and the activations are emitted as A_MULS packets - carrying the next layer's Conv-Pattern (or 16′b0 for 1×1/FC) - and written to L1 as the next layer input stream.

**Table 4. VGG-19 convolution layers specification**

| No. | Image Tensor (X, Y, C) | Filter Tensor (R, S, C, $N_F$) | | |
|---|---|---|---|---|
| 1.1 | 224, 224, 3 | 3, 3, 3, 64 | | |
| 1.2 | 224, 224, 64 | 3, 3, 64, 64 | | |
| 2.1 | 112, 112, 64 | 3, 3, 64, 128 | | |
| 2.2 | 112,112, 128 | 3, 3, 128, 128 | | |
| 3.1 | 56, 56, 128 | 3, 3, 128, 256 | Stride = 1 | Pad = 1 |
| 3.2 | 56, 56, 256 | 3, 3, 256, 256 | | |
| 3.3 | 56, 56, 256 | 3, 3, 256, 256 | | |
| 3.4 | 56, 56, 256 | 3, 3, 256, 256 | | |
| 4.1 | 28, 28, 256 | 3, 3, 256, 512 | | |
| 4.2 | 28, 28, 512 | 3, 3, 512, 512 | | |
| 4.3 | 28, 28, 512 | 3, 3, 512, 512 | | |
| 4.4 | 28, 28, 512 | 3, 3, 512, 512 | | |
| 5.1 | 14, 14, 512 | 3, 3, 512, 512 | | |
| 5.2 | 14, 14, 512 | 3, 3, 512, 512 | | |
| 5.3 | 14, 14, 512 | 3, 3, 512, 512 | | |
| 5.4 | 14, 14, 512 | 3, 3, 512, 512 | | |

*B. Evaluation Results*

    *(a) Breakdown of message-driven streaming behavior:* ***Figure 6*** depicts how the message-driven pipeline behaves in aggregate. On the left (***Figure 6(a)***), 97.85% of all messages are generated on-chip, while only 2.13% are host-injected weight messages and a negligible 0.02% are image messages. This is exactly what the deterministic schedule aims for: keep weights stationary within a fold, forward activations, and let intermediate results self-propagate - so the host barely participates after priming. On the right (***Figure***

**Table 5. Bandwidth specifications of different A)** PCIe Configurations **[38] B)** Off-Chip Memory Technologies **[39]**

| A) | |
|---|---|
| PCIe (Generation/Lanes) | Bandwidth (GB/s) |
| 1.0 | x1, x4, x8, x16 | 0.25, 1, 2, 4 |
| 2.0 | x1, x4, x8, x16 | 0.5, 2, 4, 8 |
| 3.0 | x1, x4, x8, x16 | 0.98, 3.94, 7.88, 15.8 |
| 4.0 | x1, x4, x8, x16 | 1.97, 7.88, 15.8, 31.5 |
| 5.0 | x1, x4, x8, x16 | 3.94, 15.8, 31.5, 63 |
| 6.0 | x1, x4, x8, x16 | 7.88, 31.5, 63.0, 126 |
| B) | |
| Memory (Type/Version) | Bandwidth (GB/s) |
| DDR/(-, 2, 3, 4, 5) | 0.05, 0.1, 0.2, 0.4, 0.8 |
| LPDDR/(-, 2, 3, 4X, 5, 5X) | 0.05, 0.13, 0.23, 0.53, 0.8, 1 |
| GDDR/(3, 5, 5X, 6, 7) | 0.33, 1.13, 1.5, 3, 4.5 |

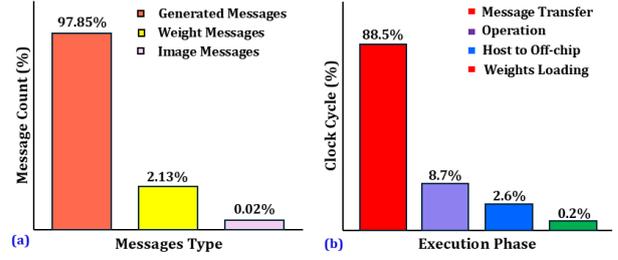

**Figure 6. Message-driven data and instruction streaming slashes host control and off-chip traffic** (>97% of messages are generated in-fabric, and ~89% of cycles are spent transferring them). **(a)** Message count (%) by type. **(b)** Clock Cycles (%) by execution phase.

*6(a)*), the cycle breakdown shows execution is transfer-dominated: message movement accounts for 88.5% of cycles, arithmetic 8.7%, host to off-chip memory transfer 2.6%, and weight loading 0.2%. In short, once primed, MAVeC runs as a transfer-bound on-chip messaging engine, slashing off-chip accesses and host interruptions while the fabric streams tasks autonomously.

    *(b) Data reuse and reduction:* ***Figure 7*** quantifies the on-chip locality exposed by the message-driven schedule. Temporal reuse (***Figure 7(a)***) grows with array size because larger fabrics hold more weights stationary per fold; it peaks in the early VGG-19 layers with wider feature maps and then tapers as spatial extents shrink. Spatial reuse (***Figure 7(b)***) shows the benefit of in-fabric multicast: each activation is fanned out once per column rather than re-fetched, so reuse scales up with array width and is again strongest in the first few layers. Spatial reduction (***Figure 7(c)***) tracks the bytes elided by the ($\Sigma_R \rightarrow \Sigma_S \rightarrow \Sigma_C$) reduction chain; it increases with channel depth and kernel size, providing another order of magnitude cut in traffic on larger arrays. Together, these mechanisms collapse external bandwidth demand and shift pressure to short on-chip transfers.

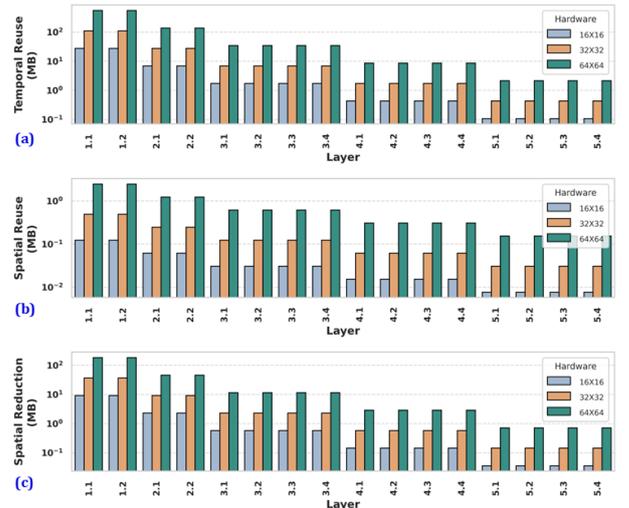

**Figure 7. Per-layer locality (log scale) under message-driven execution. (a)** Temporal reuse from stationary filter folds. **(b)** Spatial reuse from in-array multicast **(c)** Spatial reduction from the ($\Sigma_R \rightarrow \Sigma_S \rightarrow \Sigma_C$) pipeline.

    *(c) Utilization, latency, and compute throughput:* ***Figure 8*** reports per-layer utilization, latency (KCCs), and compute rate across three fabric sizes. Utilization (***Figure***

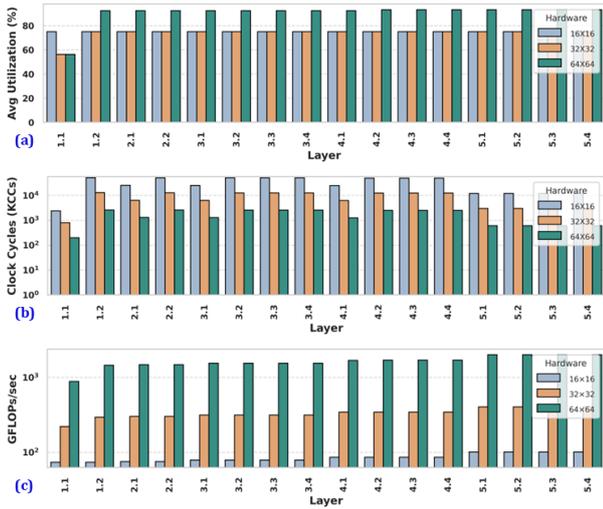

**Figure 8. Per-layer performance profiling under message-driven execution. (a)** Average SiteO Utilization (%) **(b)** Latency in Kilo Clock Cycles (log scale) **(c)** Compute throughput in GFLOPs/s (log scale)

*8(a)*) is consistently high and improves with array size: ~70–80% for 16x16, low-to-mid-80s for 32x32, and ~88–92% for 64x64. Minor dips occur when partial sum fold merges (UPDATE/A_ADD/A_ADDS), but the schedule quickly refills the pipeline. Latency (*Figure 8(b)*) falls monotonically with fabric size; more active SiteOs and short on-chip forwards reduce KCCs by roughly an order of magnitude going from 16x16 to 64x64, with the largest gains in the wide feature-map layers. Compute throughput (*Figure 8(c)*) scales nearly linearly with array dimension: the 16x16 grid sustains relatively lower GFLOP/s, 32x32 roughly doubles that, and 64×64 exceeds a teraflop for most layers. Layer-to-layer variations track problem size and mapping constructs rather than control overhead, confirming that the message-driven pipeline keeps the fabric busy.

*(d) Host link and DRAM sensitivity: Figure 9* stresses the two external knobs that could throttle a message-driven chip: the host link and the off-chip memory. In *Figure 9(a)*, throughput scales almost linearly with PCIe generation and lane count, reaching ~12 KIPS at Gen6x16. Older/leaner links (e.g., Gen3x4) leave noticeable headroom and would bottleneck execution phases with frequent host interaction (e.g., initial programming bursts or weight loading), whereas Gen5/6x8-x16 lanes are ample once the on-chip stream is primed. In contrast, sweeping DRAM families (*Figure 9(b)*) has negligible effect (~11.2–12.0 KIPS across DDR/LPDDR/GDDR) because >97% of packets are generated in-fabric, and the on-chip reuse/reduction suppress off-chip traffic.

## V. CONCLUSION

This work presented a deterministic data-instruction co-streaming framework that turns inference into an autonomous on-chip stream, eliminating host-orchestrated load/execute/store phases and making layer handoff implicit. On VGG-19, the design exhibits dominant on-chip message generation (>97%) and transfer-bound execution (~89% of cycles), with the 64x64 array sustaining 88-92% utilization and >1 TFLOP/s on most layers, with high temporal reuse, strong spatial reuse, and effective staged reductions. System throughput peaks around 12 KIPS with PCIe Gen6x16 and is largely insensitive to DRAM family. Our key distinction is the unification of mapping, scheduling, and control in a messaging-based execution model, which lets the fabric self-sequence within and across layers without host supervision.

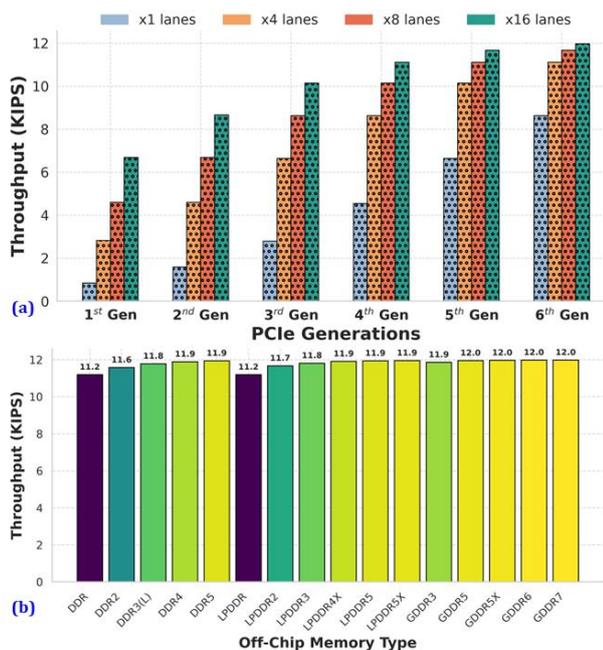

**Figure 9. I/O sensitivity under message-driven execution. (a)** Throughput vs PCIe generation and lane count. **(b)** Throughput vs off-chip memory family (DDR/LPDDR/GDDR).